# pH dependent surface enhanced Raman study of Phe + Ag Complex and DFT calculations for spectral analysis

Animesh K. Ojha[1*], Achintya Singha[1], Swagata Dasgupta[2], Ranjan K. Singh[3]
and Anushree Roy[1**]

[1]*Department of Physics and Meteorology, Indian Institute of Technology Kharagpur, 721302, India*
[2]*Department of Chemistry, Indian Institute of Technology Kharagpur 721302, India*
[3]*Department of Physics, Banaras Hindu University, Varanasi 221005, India*

___

**Abstract**

Surface enhanced Raman spectra of Phenylalanine (Phe) in Ag colloidal solution have been recorded for Phe solutions of different pH. Spectral line-shape analyses of the enhanced modes, at ~1005, 1380 and 1582 $cm^{-1}$, between pH 4.5 and 10.5, have been carried out. The variation of spectral line-width with pH reveals two possible mechanisms in solution: (i) the fluctuation of pH in microscopic volume in an overall uniform pH solution and/or (ii) the motional narrowing caused by the intermolecular ionic interaction. We suggest that different charge states of the reference molecule are responsible for the observed bond softening with decrease in pH. The observed 'Raman shift' and the 'Raman activity' of the vibrational modes with maximum enhancement have been explained by carrying out DFT calculations.

___

*\*E-mail address:* animesh_r1776@rediffmail.com
\*\*anushree@phy.iitkgp.ernet.in

## 1. Introduction

Spectroscopic techniques are useful for analytical investigations in general. For a molecule, the analysis of spectral profiles of its characteristic Raman modes allows us to obtain its static and dynamic properties in liquids and solids. For example, the position and width of a Raman mode are related to the structure and bonding of the molecule, and also to the interaction of the molecule with its environment. On the other hand, the intensity of a spectral feature carries rich information about the concentration of molecular species, polarizability of the molecule and population density of the corresponding vibrational mode.

Enhancement in Raman intensities can be achieved by bringing the molecules in contact with rough metal surfaces [1-2]. The enhancement in intensity is also observed when a molecule is adsorbed on metal particles of few tens of nanometer in size i.e. whose dimension is of the order of the wavelength of the incident light. Surface Enhanced Raman Scattering (SERS) is generally explained on the basis of two different physical causes : (i) the high local optical field at the hot sites, where the molecules are adsorbed on metal particles, due to collective excitation of conduction electrons in the small metallic structure and (ii) metal-molecule charge-transfer electronic transition. In the latter case, the resonance of electronic energy levels of the molecule and the metal atom is responsible for the enhancement. The electromagnetic effect provides an enhancement by a factor of $\sim 10^6$ for the Raman signal of the molecules adsorbed on the metal surface. The charge-transfer effect, on the other hand, provides an enhancement by a factor of $10^2$ for the chemisorbed molecules. The SERS technique has been extensively used to study different biomolecules [2-4]. In the SERS process, in addition to the enhancement in intensity of the vibrational modes, a significant frequency shift and change in Raman line-width compared to those in the free molecules are also observed [5]. To understand the time scale involved in the SERS process, time domain measurements of vibrational life-time for adsorbates on single crystal surface have been reported using picosecond vibrational spectroscopy [6]. In ref. [7-8], theoretical models have been proposed to explain the origin of the above mentioned chemical as well as electromagnetic enhancement factors.

Recently, we have reported a comparative study [9] on metal-amino acid interactions in Tyr-Ag and Phe-Ag complexes through pH dependent SERS measurements. It is to be noted that vibrational spectroscopy may access the picosecond time window, and thus, allow the investigation of collective relaxation processes. Band shape analysis of SERS modes may help in understanding the static and dynamic interactions of adsorbed molecules with Ag colloids. In the present study, a systematic analysis of the pH dependent change in line profile parameters of SERS modes for Phe-Ag amino acid-metal complex has been carried out. The optimized ground state geometry and corresponding Raman shift and Raman activity (RA) for different vibrational modes have also been calculated using Density Functional Theory (DFT) with CEP-31G basis sets using the Gaussian 98 package [10]. This basis set consists of effective core pseudo-potentials in conjunction with a double $\zeta$ contraction for valence electrons. The calculated frequencies have been uniformly scaled using a factor 0.998 for the CEP-31G basis set. The DFT with CEP-31G basis set, yields results, which are closer to the experimental values [11-12]. The results of our calculations are used to explain the profile parameters of SERS modes of Phe+Ag complex in its most favorable configuration.

## 2. Experimental and Results

Silver nitrate ($AgNO_3$), sodium borohydride ($NaBH_4$), sodium hydroxide (NaOH), hydrochloric acid (HCl) of analytical reagent grade and amino acids procured from (SRL,

India) were used to prepare the Ag colloidal solution and to maintain the pH of the solutions. The colloidal silver solution was prepared in deionized water following the method described by Creighton *et al.* [13]. The details of our sample preparation procedure are available in Ref. [9]. A $10^{-3}$ M solution of Phe was prepared in deionized water. In the sol, Ag particles are in the ionic $Ag^+$ state because of its low oxidation potential. The pH of this solution was adjusted by using 1 M HCl and 1 M NaOH. For SERS measurements, Phe solutions of different pH were added to the Ag sol (of pH value 7.86). For all experiments, the volume ratio of Ag sol to amino acid solution was maintained at 9:1. Here we would like to point out that the final pH of the mixed sol did not stabilize even after 15–20 min.

SERS spectra were measured in back-scattering geometry using a 488 nm Argon ion laser as an excitation source. The spectrometer was equipped with a 1200 grooves/mm holographic grating, a holographic super-notch filter, and a Peltier cooled CCD detector. The data acquisition time for each spectrum was 120 s. The expected spectral resolution is around 0.5 $cm^{-1}$ in the present set up.

The experimentally recorded SERS spectra (baseline subtracted) in the region 1500-3200 $cm^{-1}$ of Phe of 0.001M at six different pH values, are shown in Fig.1. The broad background due to the vibrational mode of water appears in the range 1500–1800 and 2800–3200 $cm^{-1}$. In order to subtract the baseline, this broad feature of water was normalized at a position where the Raman band was absent and then subtracted from the net SERS spectrum. All spectra in Fig. 1 have three prominent features at ~1005 ($v_1$), ~1380 ($v_2$) and ~1582 ($v_3$) $cm^{-1}$ corresponding to ring breathing of Phe and vibrational modes of $COO^-$ and $NH_3^+$, respectively. It is to be noted that these peaks are not observed in the Raman spectrum of Phe solution alone (without Ag colloids) of the same concentration (dotted line in Fig. 1). A detailed study on the variation in SERS intensities of these modes with the pH of Phe has been reported in Ref. [9]. For quantitative analysis of other spectral parameters, Raman shift and line-width, the recorded spectra for Phe in Ag sol at different pH were fitted using a non-linear curve fitting program (Spectra Calc). The curve fitting has been carried out considering the band as a mixture of Lorentzian and Gaussian, which is essentially as good as a Voigt profile. To check the uniqueness of the fitting parameters, each spectrum has been fitted with different reasonable initial guesses and each time we have obtained the same fitted profiles and fitting parameters. The experimental as well as analyzed spectra at different pH values corresponding to $\nu_1$ and $\nu_2$ bands are separately presented in Fig. 2 and Fig. 3, respectively.

### 3. pH dependent shift in peak positions and line-widths of SERS bands

The variation in line-width and peak positions of all the three SERS bands; $\nu_1$, $\nu_2$ and $\nu_3$, as obtained from the non-linear curve fitting, are shown in Table 1. In Figs 4 (a) and (b), we have presented the variation of the above parameters for $\nu_2$ and $\nu_3$ bands at different values of pH. The changes for the $\nu_1$ band lie within our experimental accuracy. It is evident that the peak positions of all the above features show a linear upward shift with an increase in pH of Phe solutions. Here we would like to point out that for the Raman spectrum of Phe at different pH (without $Ag^+$ colloids in solution) we did not observe such a systematic change in the Raman shift.

For different pH values, the Phe molecules in solution are in different charge states [9]. The structural features of the Phe molecules with $NH_3^+$ and COOH groups is found in highly acidic solutions below $pK_1$ of 2.5. Above the pH value of 2.58, COOH becosmes $COO^-$, however, all $NH_3^+$ changes to $NH_2$ only above the pH value of 9.24 ($pK_2$ : 9.24). From the observed variation in peak positions of SERS bands with the pH of the Phe, shown in Fig. 4, we can conclude that the anionic and the cationic forms of Phe in Ag colloidal solution affect the vibrational frequencies of the oscillators. The upward shift in frequencies of the oscillators can be explained in terms of the increase in charge transfer between metal particle and reference molecule as a result of different charge states of the molecule for different pH. The electronic charge transfer increases the vibrational force constant and hence, the frequency of the oscillators [14]. Here we refer to our earlier work in Ref. [9], where we have shown the increase in intensities of the SERS bands at 1005 and 1380 cm$^{-1}$ between pH 4.5 and 9.5 of Phe, which is a clear signature of an increase in charge transfer between the reference molecule and Ag$^+$ in the colloidal solution between these pH values.

The line-width vs. pH plot of the 1380 cm$^{-1}$ band of COO$^-$, shown in Fig. 4(a), depicts a Gaussian type variation with a maximum at pH (pH$_0$) at 7.57 (obtained after fitting the data points using a Gaussian function). It decreases both on increasing and decreasing the pH value of Phe. However, the line-width of the band at 1582 cm$^{-1}$ due to NH$_3^+$, linearly decreases with pH between 3.5 and 10.5. We make an attempt to explain the above variation in line-widths with the following arguments. The phenomenon of the occurrence of a maximum in the FWHM for noncritical concentration fluctuations in binary fluids has been observed earlier [15-16]. In analogy with this concentration fluctuation model, we propose that the random motion of solute molecules throughout the volume of the colloidal solution causes the pH to fluctuate around a mean value of pH$_0$ in a Gaussian manner. Indeed, as mentioned in Section 2, we have observed in our experiment that the pH of Phe+Ag complex did not stabilize even after 15-20 minutes (for more details see Ref. [9]). However, the above model is expected to predict a wider variation in line-width with pH compared to what we have observed in Fig. 4(a). Furthermore, with an increase in pH, the transition from the cationic to zwitterionic and zwitterionic to anionic form of the reference molecule results in its higher mobility due to the decrease in intermolecular interaction between $COO^-$ and $NH_3^+$ groups [15]. Hence, the vibrational relaxation time of the molecule is expected to decrease for higher pH values. Thus, one can expect an increase in line-width with an increase in pH. If the excitation takes place Concurrently, with an increase in pH, due to decrease of solvation while going from $-NH_3^+$ to $-NH_2$, an 'apparent' increase in the viscosity of the solution takes place [17]. This, in turn, decreases the mobility of the reference molecule and consequently results in a decrease in its spectral line-width. The competition between the above two processes determines the net spectral width. The effect of viscosity, as expected, dominates for the SERS band of $NH_3^+$. Thus, in our case, it shows a linear decrease in vibrational line-width with increase in pH. On the other hand, for the vibrational mode of $COO^-$, the effect of intermolecular interaction dominates over the effect of viscosity in the lower pH range and hence we observe an increase in line-width with initial increase in pH of Phe.

## 4. DFT calculations for optimized geometry of Phe+Ag complex and its vibrational wavenumbers

The basic objective of DFT calculations in this study is two fold. Firstly, to calculate the optimized geometry of Phe and metal complex and secondly, to calculate the vibrational (Raman) spectral parameters, particularly Raman shift and RA, of the resulting structure. To explain the experimental results on the basis of theoretical calculations we used different functions and basis sets. The result has been used to explain the experimental observations related to characteristics of the spectral features of Phe in Ag solution for maximum SERS.

We have seen in Ref. [9], that the zwitterionic form of Phe in $[Phe+Ag]^+$ complex exhibits maximum SERS activity. To find the most favorable site for the Phe molecule of this form for the attachment of $Ag^+$, the calculations have been carried out in four possible configurations as shown in Fig 5. The theoretical calculation yields a number of putative binding geometries for [Phe+Ag]$^+$ for which the free energy is very close. Fig. 5(a) shows the structure of the zwitterionic Phe molecule in the minimum energy configuration. To understand SERS spectra, the Ag atom is kept either near the N atom [Fig. 5(b)], O atom [Fig. 5(c)], or the benzene ring [Fig. 5(d)] of the reference molecule. The fourth configuration involves a tridentate binding of the N, O, and the π-electron cloud of the aromatic ring [Fig. 5(e)]. The optimized energies corresponding to the isolated Phe, [Fig. 5(a)] and to each configuration in Figs. 5(b) -5(e) are -2.698, -6.675, -6.674, -6.673 and –6.676 KeV, respectively. In terms of the minimum energy of the [Phe+Ag]$^+$ complex, the configuration shown in Fig. 5(e) seems to be the most probable. However, the possibility of other three configurations cannot be completely ruled out. Next, all four resulting metal-amino acid ligand complexes, shown in Fig. 5 (b) – Fig. 5(e), are used to calculate the vibrational wavenumbers and Raman activities for the three modes, $\nu_1$, $\nu_2$, and $\nu_3$, discussed in this article, and are given in Table 2. The enhancement of a Raman mode can be looked at through its Raman activity based on the change in the amplitude of polarizability. In our calculation, Raman activity of the SERS band at ~ 1005.0 cm$^{-1}$ ($\nu_1$) is enhanced (RA 34 → RA 41) in two configurations as shown in Fig. 5(b) and 5(c); whereas, there is no enhancement for the complex shown in Fig. 5(d). For the configuration shown in Fig. 5(e), we have observed the maximum enhancement (RA 34 → RA 42.8) of the same mode. In case of the SERS band at ~1379.8 cm$^{-1}$ ($\nu_2$), the enhancement is maximum for the complex shown in Fig. 5(b). It is also appreciable for the complex shown in Fig. 5(e). The SERS band at 1582 cm$^{-1}$ ($\nu_3$) also exhibits maximum enhancement (RA 83.4 → RA 190.0) for the configuration shown in Fig. 5(e). In other words, the complex shown in Fig. 5(e) exhibits maximum enhancement for all three modes, as observed by us experimentally in Ref. [9] and as shown in Fig. 1. In addition, the Raman shifts of the above modes in this particular configuration are found to be very close to the observed shifts in the spectral lines. The DFT configuration of [Phe + Ag]$^+$ complexes, in conjunction with our experimental observations explains a long standing query as to why a strong SERS band for $NH_3^+$ is observed for [Phe+Ag]$^+$ complex. Thus, we confirm that the complex shown in Fig. 5(e) is the most favorable configuration for Phe-Ag ligand-metal complex.

## 5. Conclusions

We have reported the enhancement and spectral profiles of three vibrational modes at ~ 1005, 1380 and 1582 cm$^{-1}$ in the SERS spectra of the Phe-Ag complex. The enhancement of these modes is highly pH dependent and exhibits a maximum at a pH value of 9.5. The variation in peak positions with pH of Phe in the colloidal solution of Ag has been explained by using the charge transfer phenomenon between the metal and the reference molecule. The phenomenon of motional narrowing due to intermolecular interaction and change in viscosity of the solution possibly explains the change in line-width of the spectra. In the colloidal solution, the energy minimized configuration for the [Phe+Ag]$^+$ complex has been looked at through DFT calculations. The optimized geometry with minimum free energy and the vibrational spectral parameters thus obtained have been compared with the experimentally observed RA and the Raman shift of different vibrational modes.
## Acknowledgments

AKO and AKS would like to thank the CSIR, India, for the financial support. AR thanks DST, India for financial assistance.

Table 1
Observed Raman shift and line-widths of the three SERS bands in the spectra of (Phe + Ag sol.) with varying pH of Phe.

| pH of Phe | Observed SERS bands of Phe | | | | | |
|---|---|---|---|---|---|---|
| | ~ 1004 cm$^{-1}$ band | | ~ 1380 cm$^{-1}$ band | | ~ 1582 cm$^{-1}$ band | |
| | Peak Position (cm$^{-1}$) | Linewidth (cm$^{-1}$) | Peak Position (cm$^{-1}$) | Linewidth (cm$^{-1}$) | Peak Position (cm$^{-1}$) | Linewidth (cm$^{-1}$) |
| 10.5 | 1004.5 | 5.53 | 1379.8 | 32.0 | | |
| 9.5  | 1005.0 | 5.01 | 1379.5 | 32.4 | 1581.6 | 12.53 |
| 8.5  | 1005.5 | 5.81 | 1379.0 | 34.3 | 1581.0 | 13.00 |
| 7.5  | 1004.7 | 5.40 | 1378.9 | 35.2 | 1580.6 | 14.67 |
| 6.5  | 1004.4 | 5.82 | 1378.4 | 34.0 | | |
| 5.5  | 1003.9 | 5.72 | 1377.5 | 32.2 | 1579.0 | 16.39 |
| 4.5  | 1003.5 | 6.05 | | | | |

Table 2
Measured and the difference of measured and calculated Raman shift (RS), and Raman Activity (R. A.) of three SERS bands in different configurations of Phe + Ag complexes. Calc. A = Phe, Calc. B = Phe + Ag atom near to N atom, Calc. C = Phe + Ag atom near to O atom, Calc. D = Phe + Ag atom near to benzene ring, E = Phe + Ag atom near to benzene ring, N and O

| | Obs. | Calc. A | | Calc. B | | Calc. C | | Calc. D | | Calc. E | |
|---|---|---|---|---|---|---|---|---|---|---|---|
| | RS cm$^{-1}$ | RS cm$^{-1}$ | RA | RS cm$^{-1}$ | RA | RS cm$^{-1}$ | RA | RS cm$^{-1}$ | RA | RS cm$^{-1}$ | RA |
| SERS $\nu_1$ | 1005.0 | +0.5 | 34.0 | -1.0 | 41.0 | -1.8 | 41.0 | -1.6 | 33.5 | -0.9 | 42.8 |
| SERS $\nu_2$ | 1379.5 | +42.5 | 4.7 | +37.5 | 62.4 | +15.5 | 38.2 | +18.5 | 3.6 | +3.0 | 58.8 |
| SERS $\nu_3$ | 1581.6 | +83.4 | 4.8 | +83.4 | 20.0 | +89.4 | 6.8 | +93.4 | 7.0 | +58.0 | 190.0 |

## FIGURE CAPTIONS

**Fig. 1:** Surface enhanced Raman spectra of Phe at six different pH in the region 500-3000 cm$^{-1}$. *, $ and + indicate the three bands, discussed as $\nu_1, \nu_2$ and $\nu_3$ in the text.

**Fig. 2:** Surface enhanced Raman spectra of ring breathing mode of Phe at different pH values. Continuous and dotted lines represent the experimental and fitted spectra, respectively.

**Fig. 3:** Same as Fig.2 except for COO$^-$ mode.

**Fig.4:** Variation in linewidth and wavenumber of the vibrational modes at (a) 1380 cm$^{-1}$ due to COO$^-$ group and (b) 1582 cm-1 due to NH$_3^+$ mode increase in pH Phe.

**Fig. 5:** Energy minimized structures of (a) isolated Phe and [Phe+Ag]+ complex for when Ag is placed (b) near the N terminal (c) near the O terminal (d) near the ring structure of the reference molecule. In (e) Ag is preferentially occupied in the cavity of the reference molecule.

**Fig. 1**

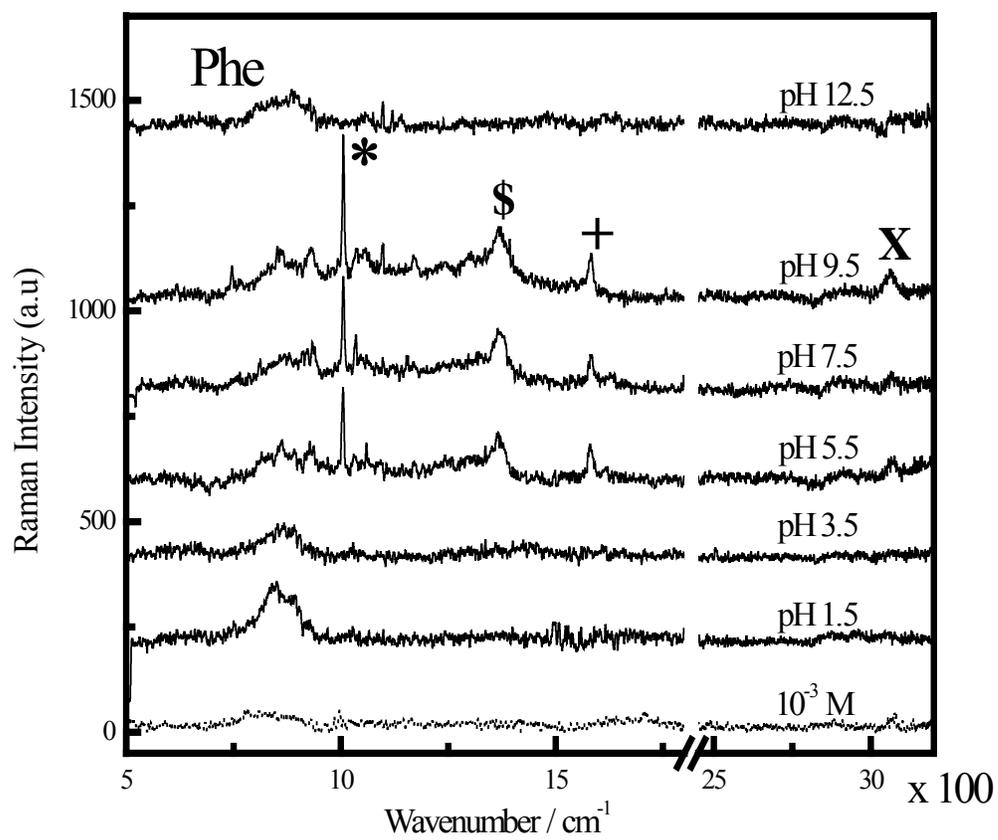

**Fig. 2**

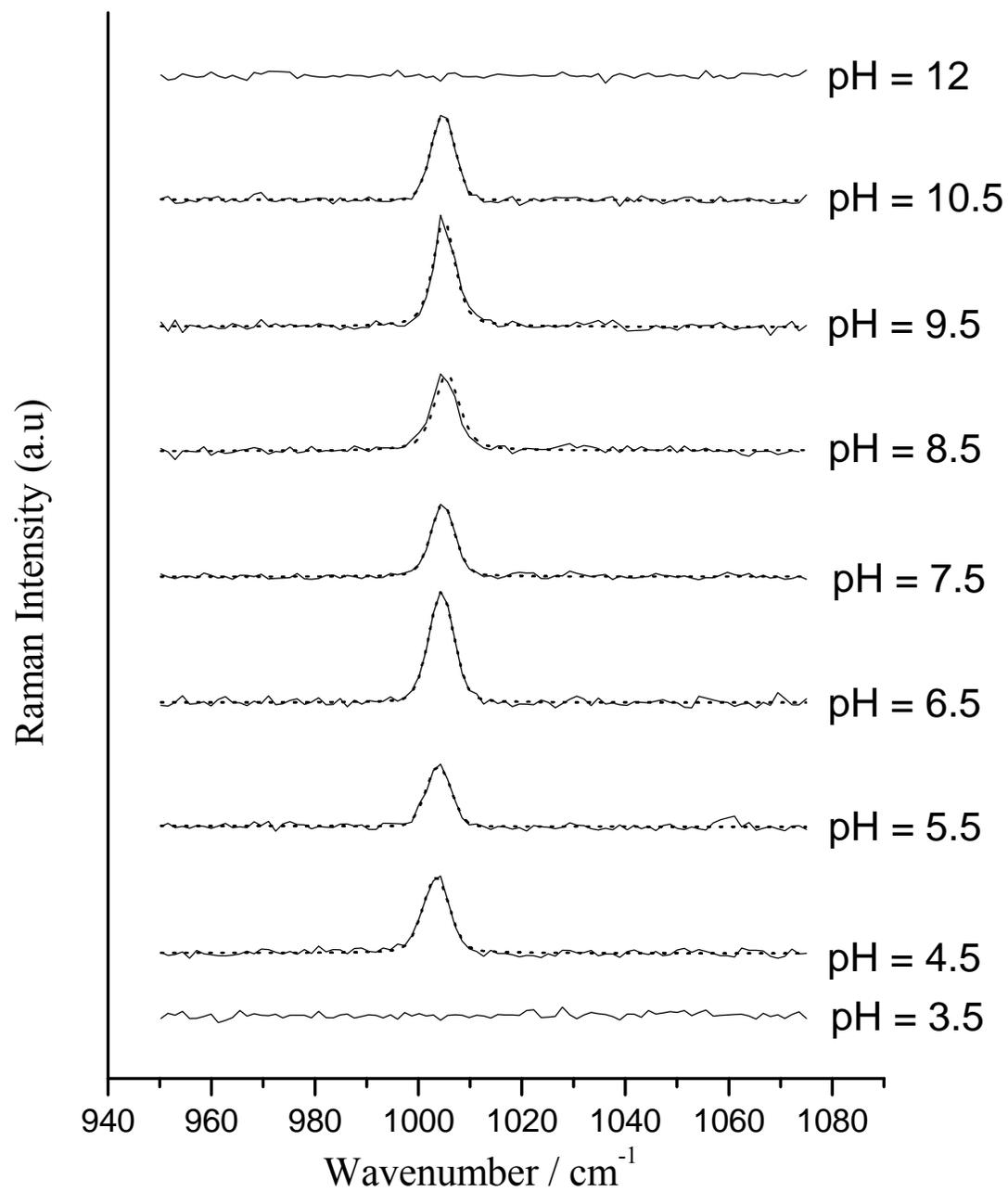

**Fig. 3**

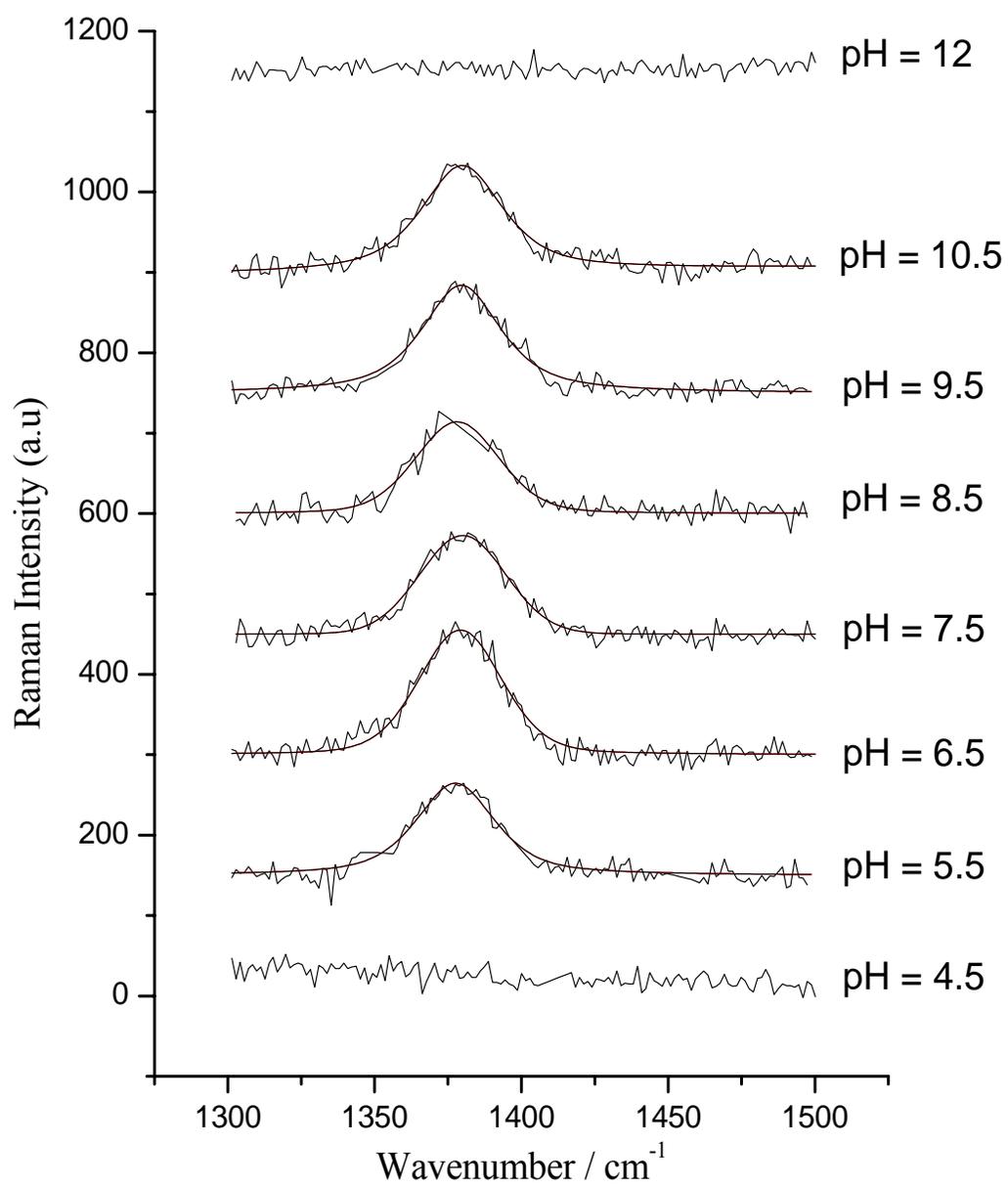

**Fig. 4**

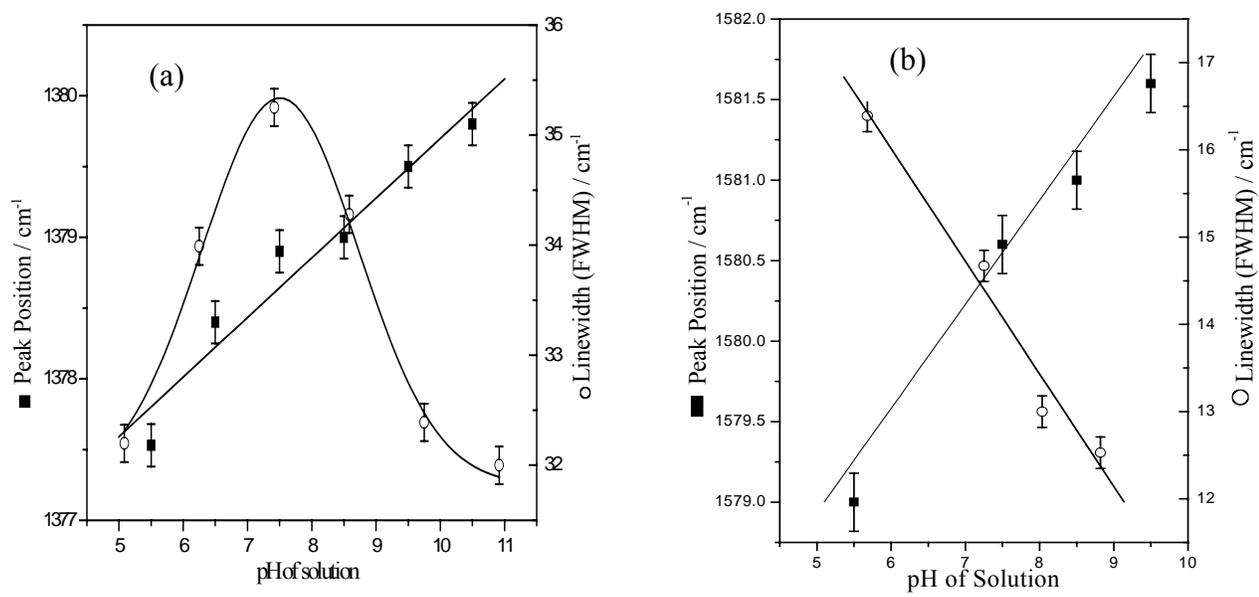

**Fig. 5**

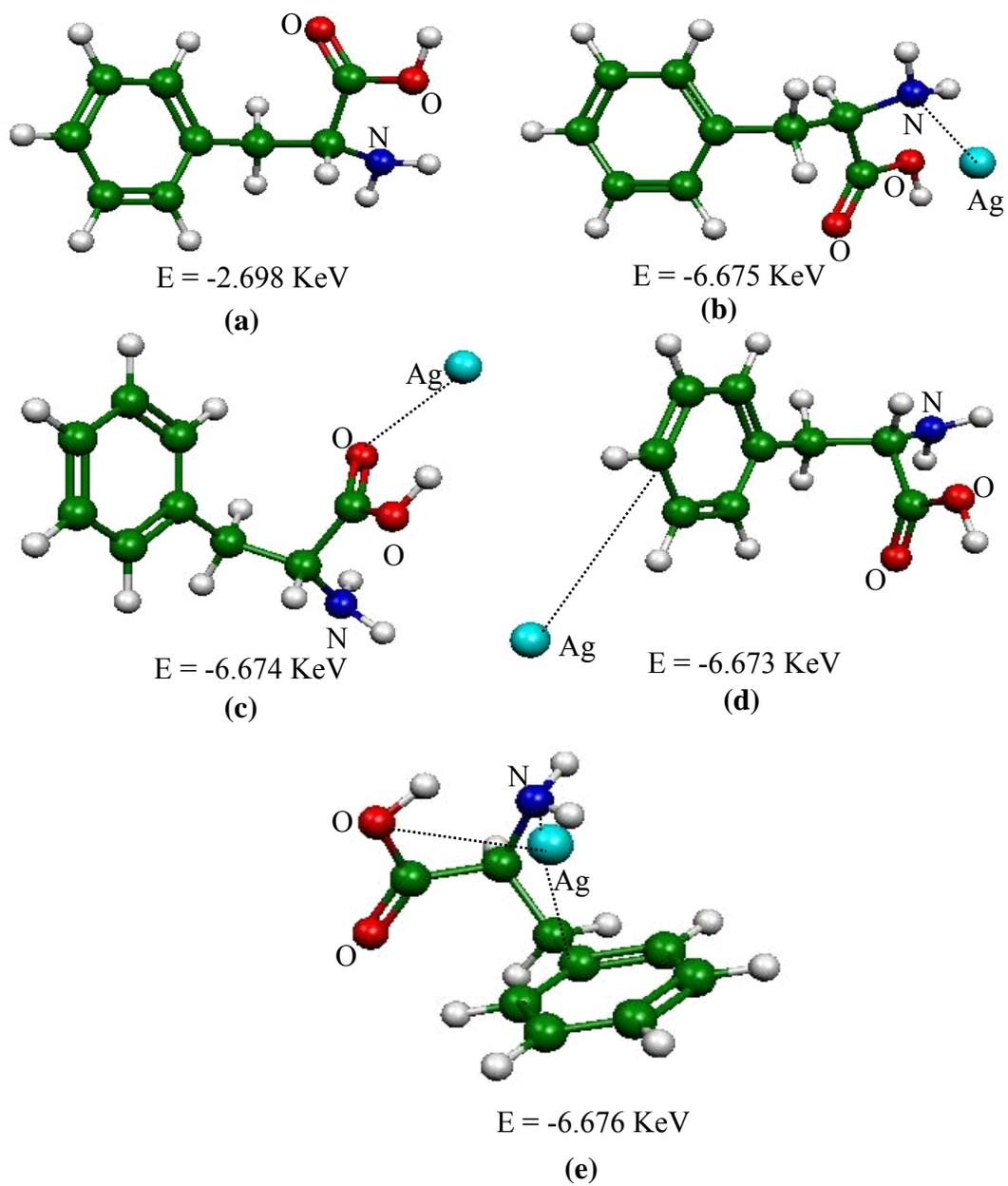

E = -2.698 KeV
(a)

E = -6.675 KeV
(b)

E = -6.674 KeV
(c)

E = -6.673 KeV
(d)

E = -6.676 KeV
(e)